\documentclass{article}
\usepackage{spconf,amsmath,epsfig,amssymb,xcolor,bbm,theorem,enumerate,graphicx}
\usepackage[linesnumbered,vlined,boxed,commentsnumbered]{algorithm2e}
\usepackage{cleveref}

\def\nset{{\mathbb{N}}}
\def\rset{\mathbb R}

\def\Zset{\mathsf{Z}}

\def\rmd{\mathrm{d}}
\def\argmin{\operatorname{Argmin}}

\def\max{\mathrm{max}}

\def\1{\mathbbm{1}}

\def\PP{\mathbb{P}} 
\def\PE{\mathbb{E}} 

\newcommand{\F}{\mathcal{F}} 
\def\pen{\mathrm{pen}} 

\newcommand{\pscal}[2]{\left\langle#1,#2\right\rangle}
\def\Id{\mathrm{I}}
\newcommand{\eqdef}{\ensuremath{\stackrel{\mathrm{def}}{=}}}

\newcommand{\kouter}{k_\mathrm{out}}
\newcommand{\kin}{k_\mathrm{in}}

\newcommand\init{\mathrm{init}}
\newcommand{\R}{\mathsf{R}}

\newcommand{\s}{\mathsf{s}}

\newcommand{\mf}{\mathsf{h}}
\newcommand{\mfint}{\mathcal{H}}
\newcommand{\hatmf}{\widehat{\mf}}
\newcommand{\hatS}{\widehat{S}}
\newcommand{\Smem}{\mathsf{S}}

\newcommand{\Sset}{\mathcal{S}}
\newcommand{\Vois}{\mathcal{V}}
\newcommand{\error}{\eta}
\newcommand{\nbrmc}{m}

\newcommand{\Prox}{\mathrm{Prox}}

\newcommand{\param}{\theta}

\newcommand{\lyap}{\operatorname{W}}
\newcommand{\batch}{\mathcal{B}}
\newcommand{\lbatch}{\mathsf{b}}
\newcommand{\pas}{\gamma}

\def\indic{\chi}

\def\eqsp{\;}

\newcommand{\ocint}[1]{\left(#1\right]}

\newcommand{\ccint}[1]{\left[#1\right]}

\newtheorem{assumption}{A\hspace{-3pt}}

\newtheorem{theorem}{Theorem}

\title{The Perturbed Prox-Preconditioned SPIDER algorithm: non-asymptotic convergence bounds } \name{G. Fort$^1$, E. Moulines$^2$, \thanks{This
      work is partially supported by the {\em Fondation Simone et Cino
        Del Duca} through the project OpSiMorE and by the ANR-19-CHIA-0002-01. Part of this work was conducted under the auspices of the Lagrange Center in Mathematics and Computer Sciences}} \address{$^1$ IMT,
    Universit\'e de Toulouse \& CNRS, F-31062 Toulouse, France.
    \\ $^2$ CMAP, Ecole Polytechnique, Route de Saclay, 91128
    Palaiseau Cedex, France.  }

\begin{document}
\ninept
\maketitle
\begin{abstract}
A novel algorithm named {\tt Perturbed} {\tt Prox-Preconditioned
  SPIDER (3P-SPIDER)} is introduced. It is a stochastic
variance-reduced proximal-gradient type algorithm built on {\tt
  Stochastic Path Integral Differential EstimatoR} (SPIDER), an
algorithm known to achieve near-optimal first-order oracle inequality
for nonconvex and nonsmooth optimization. Compared to the vanilla
prox-SPIDER, \texttt{3P-SPIDER} uses preconditioned gradient
estimators. Preconditioning can either be applied "explicitly" to a
gradient estimator or be introduced "implicitly" as in applications to
the EM algorithm.  \texttt{3P-SPIDER} also assumes that the
preconditioned gradients may (possibly) be not known in closed
analytical form and therefore must be approximated which adds an
additional degree of perturbation. Studying the convergence in
expectation, we show that \texttt{3P-SPIDER} achieves a near-optimal
oracle inequality $O(n^{1/2} /\epsilon)$ where $n$ is the number of
observations and $\epsilon$ the target precision even when the
gradient is estimated by Monte Carlo methods. We illustrate the
algorithm on an application to the minimization of a penalized
empirical loss.
\end{abstract}
\begin{keywords}
  Statistical Learning, Large Scale Learning, Variance reduced Stochastic gradient, Finite sum optimization, Control Variates.
\end{keywords}
\section{Introduction}
\label{sec:intro}
Consider the following composite, nonconvex, and possibly nonsmooth optimization problem
\begin{equation}
\label{eq:problem-0}
\argmin_{s \in \Sset} \{ \lyap(s) + g(s) \} \eqsp,
\end{equation}
where $\Sset$ is a closed convex subset of $\rset^q$,
$\lyap:\mathcal{V} \to \rset$ is a smooth function defined on a
neighborhood $\mathcal{V}$ of $\Sset$ and $g: \Sset \to
\ocint{-\infty, + \infty}$ is a proper lower semi-continuous convex
function (with an easy to compute proximal term). This paper addresses
the case when $\lyap$ has a finite-sum structure
\begin{equation}
\label{eq:problem-01}
\lyap(s) = \frac{1}{n} \sum_{i=1}^n \lyap_i(s) \eqsp,
\end{equation}
and the optimization problem \eqref{eq:problem-0} is solved by a {\em
  preconditioned}-gradient based algorithm.  Optimization problems
\eqref{eq:problem-0}-\eqref{eq:problem-01} often arise in machine
learning. In such case, $n$ is the number of examples which is
typically very large, $\lyap_i$ is the loss function associated to
example $\# i$, and $g$ is a non-smooth regularization term,
\textit{e.g.}  $\ell_1$ norm, Elastic net, etc.  The preconditioning
setting may naturally arise "implicitly", for example in the
stochastic finite-sum version of the Expectation-Maximization (EM)
algorithm in the exponential family, which was the main motivation for
this work; see \Cref{sec:logistic} and the companion paper
\cite{fort:moulines:SSP21:algorithm}.

Minimization problems \eqref{eq:problem-0} and \eqref{eq:problem-01}
cover a broad range of applications in machine learning, statistics,
and signal processing; see \cite{bottou2018optimization}.
State-of-the art methods to solve these problems rely on stochastic
optimization approaches
\cite{johnson2013accelerating,schmidt2017minimizing}. In the nonconvex
case, while numerical algorithms for solving the noncomposite setting
(\textit{i.e.} $g=0$), are well-developed and have received
significant attention \cite{allen2017natasha,allen2017neon2}, methods
for composite optimization remain scarce
\cite{reddi2016proximal,wang:etal:nips:2019}. The authors in
\cite{reddi2016proximal} proposes and studies a non-composite
finite-sum problem using {\tt SVRG} estimator from
\cite{johnson:zhang:2013}. This method is extended to the composite
setting by applying the proximal operator of $g$ as in the
proximal-gradient scheme (see
\cite{beck:teboulle:2010,combettes:pesquet:2011,parikh:boyd:2013} for
literature review on the proximal-gradient algorithm). This technique
is based on gradients and does not use preconditioning. This scheme
has been later improved with {\tt SPIDER}, where the gradient control
variate is sequentially updated to improve the estimation accuracy:
{\tt SPIDER} is known to achieve near optimal oracle complexity in
nonconvex
optimization~\cite{nguyen:liu:etal:2017,fang:etal:2018,wang:etal:nips:2019}.

This paper analyzes the {\tt 3P-SPIDER} algorithm designed to solve
\begin{equation}
\label{eq:problem}
s \in \Sset: \qquad 0 \in \nabla W(s) + \partial g(s) \eqsp,
\end{equation}
by combining \textit{(i)} a variance-reduced preconditioned-gradient
algorithm designed for the finite sum setting, and \textit{(ii)} a
proximal step to take into account the (non smooth) regularizer $g$.
Furthermore {\tt 3P-SPIDER} covers the case when the preconditioned
gradient - of the form $n^{-1} \sum_{i=1}^n \mf_i(s)$, see
A\ref{hyp:precond} below - is not computable in a closed-form and
is approximated: both the cases of an approximation of the full sum
over $n$ terms by a sum over a random subsample, and an approximation
of the functions $\mf_i$'s are considered. {\tt 3P-SPIDER} was
introduced in \cite{fort:moulines:SSP21:algorithm} for a specific
application to large scale learning solved by a Expectation
Maximization-based optimization method. The main contribution of this
paper is to provide explicit control of the convergence in expectation
of {\tt 3P-SPIDER} and deduce complexity bounds in terms of the design
parameters of this algorithm. A comparison to the state of the art
gradient-based methods in terms of complexity bounds, is also provided
in the case the quantities $\mf_i$ are expectations and are
approximated by a Monte Carlo integration: it is established that the
number of Monte Carlo samples can be chosen so that {\tt 3P-SPIDER}
reaches the oracle complexity bounds corresponding to the case where
the $\mf_i(s)$'s are known in closed form.

{\bf Notations.} $\rset_+^\star$ and $\nset^\star$ denote respectively
(resp.) the positive real line and the positive integers. For $n \in
\nset^\star$, set $[n]^\star \eqdef \{1, \cdots, n\}$ and $[n] \eqdef
\{0, \cdots, n\}$.  For $x \in \rset$, $\lceil x \rceil$ is the
nearest integer greater than or equal to $x$. Vectors are
column-vectors; for $a,b$ in $\rset^\ell$, $\pscal{a}{b}$ denotes the
Euclidean scalar product, and $\|a\|$ the associated norm. For a
matrix $A$, we denote by $A^T$ and $A^{-1}$ resp. its transpose and
its inverse. $\Id_d$ is the $d \times d$ identity matrix.  The random
variables are defined on a probability space $(\Omega, \mathcal{A},
\PP)$; $\PE$ denotes the associated expectation. For random variables
$U$ and a sigma-field $\F$, $\PE[U \vert \F]$ is the conditional
expectation of $U$ given $\F$. For a smooth function $f$, $\nabla f$
is the gradient of $f$. For a proper lower semi-continuous convex
function $g$ and $x$ in its (assumed) non-empty domain, $\partial
g(x)$ is the subdifferential of $g$ at $x$.

\section{The 3P-SPIDER algorithm}
The optimization problem at hand is the problem \eqref{eq:problem} in
the case when
\begin{assumption} \label{hyp:Prox}
$\Sset$ is a closed convex subset of $\rset^q$. \\ $\lyap: \Vois \to
  \rset$ is a continuously differentiable function on $\Vois$, an open
  neighborhood of $\Sset$. Its gradient $\nabla \lyap$ is globally
  Lipschitz-continuous on $\Sset$ with Lipschitz constant $L_{\dot
    \lyap}$.  \\ $g: \Sset \to \ocint{-\infty, + \infty}$ is a proper
  lower semi-continuous convex function.
\end{assumption}
We consider a gradient approach for solving \eqref{eq:problem} and
allow the use of a preconditioning matrix $B(s)$ which may depend on
the current value of the parameter $s$.  We assume that
\begin{assumption}\label{hyp:precond}
For any $s \in \Sset$, $B(s)$ is a $q \times q$ positive
definite matrix and there exist $0 < v_{\min} \leq v_{\max}$ such that
for any $s \in \Sset$, the spectrum of $B(s)$ is in $\ccint{v_{\min},
  v_{\max}}$. \\ For all $i \in [n]^\star$, there exists a globally
Lipschitz function $\mf_i: \Sset \to \rset^q$, with constant $L_i$,
such that
\[
- B^{-1}(s) \, \nabla \lyap(s) = \frac{1}{n} \sum_{i=1}^n \mf_i(s)
\eqsp.
\]
\end{assumption}
We introduce
a weighted proximal operator: for a $q \times q$ positive definite
matrix $B$, define for any $s \in \rset^q$ and $\pas>0$,
\[
\Prox_{B, \pas g}(s) \eqdef \argmin_{s' \in \Sset} \left\{ \pas g(s')
+ \frac{1}{2} (s'-s)^T B (s'-s) \right\} \eqsp.
\]
Set $\mf(s) \eqdef n^{-1} \sum_{i=1}^n \mf_i(s)$. Under
A\ref{hyp:Prox} and A\ref{hyp:precond}, for any $s,s' \in \Sset$ and $\pas > 0$,
$\Prox_{B(s), \pas g}(s')$ exists and is unique and, since  $s'= \Prox_{B(s),\pas g}(s + \pas \mf(s))$ if and only if $0 \in \partial g(s) + B(s) (s'-s- \pas \mf(s))$, we obtain 
\begin{multline}
  \{s \in \Sset: \Prox_{B(s),\pas g}(s + \pas \mf(s)) =s \} \\ = \{ s
  \in \Sset: 0 \in  \nabla \lyap(s) + \partial g(s)\} \eqsp. \label{eq:fixedpoint}
\end{multline}
This property implies that the solutions of \eqref{eq:problem} are the
roots of the function $s \mapsto \Prox_{B(s), \pas g}(s+ \pas \mf(s))-s$
restricted to $\Sset$, whatever $\pas >0$.

For any minibatch $\batch$ of size $\lbatch$, sampled at random from
$[n^\star]$ - with or without replacement, we have (see
e.g. \cite[Lemma 4]{fort:moulines:wai:2020})
\[
\mf(s) = \frac{1}{\lbatch} \PE\left[ \sum_{i \in \batch} \mf_i(s)
  \right] \eqsp,
\]
thus implying that in the finite-sum setting, the preconditioned
gradient $-B^{-1}(s) \nabla \lyap(s)$ can be approximated by a sum
with $\lbatch$ terms where the indexes of summation $i$ are sampled
randomly: $\lbatch^{-1} \sum_{i \in \batch} \mf_i(s)$. Therefore, a
natural extension of the Proximal-Gradient algorithm to the finite-sum
setting would define a sequence $\{\hatS_k, k \geq 0 \}$ by
\[
\hatS_{k+1} = \Prox_{B(\hatS_k), \pas_{k+1} g} \left( \hatS_k +
\frac{\pas_{k+1}}{\lbatch} \sum_{i \in \batch_{k+1}} \mf_i(
\hatS_k)\right)
\]
where $\{\pas_k, k \geq 0\}$ is a positive deterministic sequence and
$\{\batch_{k+1}, k \geq 0\}$ is a sequence of minibatches of size
$\lbatch$ sampled at random from $[n]^\star$. {\tt 3P-SPIDER} reduces
the variance of this stochastic perturbation by the construction of a
control variate, which is defined as an approximation of
$\mf(\hatS_k)$ correlated with the random variable $\lbatch^{-1}
\sum_{i \in \batch_{k+1}} \mf_i(\hatS_k)$. This control variate is
refreshed at each so-called {\em outer loop}, indexed by $t$ in
Algorithm~\ref{algo:3PSPIDER}; and then evolves along the {\em inner
  loops}, indexed by $k$. Finally, {\tt 3P SPIDER} allows
approximations on the computation of $\mf_i(\hatS_{t,k})$,
approximations denoted by $\hatmf_i^{t,k}$.

The algorithm is given in Algorithm~\ref{algo:3PSPIDER}.  At the start
of each outer loop $\# t$, the control variate $\Smem_{t,0}$ is
initialized (see Lines~\ref{eq:SA:reset:0} and \ref{eq:SA:reset:1}) in
order to approximate $\mf(\hatS_{t,-1})$; a natural idea is to choose
$\mathcal{E}_t=0$.  Nevertheless, the computational cost is important
since it involves a sum over $n$ terms, and this full sum can be
substituted with a sum having $\lbatch' \ll n$ indices defined by a
minibatch $\batch_{t,0}'$ sampled at random from $[n]^\star$; in that
case, $\mathcal{E}_t \neq 0$. The control variate is modified at each
inner loop $\# k$ (see Line~\ref{eq:SA:update:Smem}): since
$\Smem_{t,0} \approx \mf(\hatS_{t,-1})$, we have $\Smem_{t,k+1}
\approx \mf(\hatS_{t,k})$ upon noting that $\Smem_{t,k+1} -
\Smem_{t,k} \approx \mf(\hatS_{t,k}) - \mf(\hatS_{t,k-1})$. The key
property is the choice of the same minibatch $\batch_{t,k+1}$ when
approximating $ \mf(\hatS_{t,k})$ and $ \mf(\hatS_{t,k-1})$: the
correlation of these quantities is the essence of the {\em control
  variate} mechanism.

\begin{algorithm}[htbp]
    \caption{The Perturbed Prox-Preconditioned SPIDER (3P-SPIDER)
      algorithm.
      \label{algo:3PSPIDER}}
   \KwData{ $\kouter, \kin \in \nset^\star$; $\hatS_\init \in \Sset$;
     $\pas_{t,0} \geq 0$, $\pas_{t,k} >0$ for $t \in [\kouter]^\star$,
     $k \in [\kin]^\star$.}  \KwResult{The 3P-SPIDER sequence
     $\{\hatS_{t,k}, t \in [\kin]^\star, k \in [\kin]\}$} $\hatS_{1,0}
   = \hatS_{1,-1} = \hatS_\init$ \; $\Smem_{1,0} = n^{-1} \sum_{i=1}^n
   \hatmf_i^{1,-1}+ \mathcal{E}_1$ \label{eq:SA:reset:0} \; \For{$t=1,
     \cdots, \kouter$ \label{eq:SA:epoch}}{ \For{$k=0,
       \ldots,\kin-1$}{Sample a mini batch $\batch_{t,k+1}$ of size
       $\lbatch$ in $[n]^\star$ \label{eq:SA:update:batch} \;
       $\Smem_{t,k+1} = \Smem_{t,k} + \lbatch^{-1} \sum_{i \in
         \batch_{t,k+1}} \left( \hatmf_i^{t,k} - \hatmf_i^{t,k-1}
       \right) $ \label{eq:SA:update:Smem} \; $\hatS_{t,k+1/2} =
       \hatS_{t,k} + \pas_{t,k+1} \Smem_{t,k+1}
       $ \label{eq:SA:updateclassical} \; $\hatS_{t,k+1} =
       \Prox_{B(\hatS_{t,k}), \pas_{t,k+1} g}\left( \hatS_{t,k+1/2}
       \right)$ \label{eq:SA:proximal} \;} $\hatS_{t+1,-1} =
     \hatS_{t,\kin}$ \; $\Smem_{t+1,0}= n^{-1} \sum_{i=1}^n
     \hatmf_i^{t+1,-1} + \mathcal{E}_{t+1}$ \label{eq:SA:reset:1} \;
     $\hatS_{t+1,-1/2} = \hatS_{t+1,-1} + \pas_{t+1,0} \Smem_{t+1,0} $
     \; $\hatS_{t+1,0} = \Prox_{B(\hatS_{t+1,-1}),
       \pas_{t+1,0}}(\hatS_{t+1,-1/2})$}
\end{algorithm}

The {\tt SPIDER} algorithm
(\cite{nguyen:liu:etal:2017,fang:etal:2018,wang:etal:nips:2019})
corresponds to the case $g=0$, $\Sset= \rset^q$, $B(s) = \Id_q$ and
$\hatmf^{t,k}_i = \mf_i(\hatS_{t,k})$ for any $i \in [n]^\star$, $t
\in [\kouter]^\star$ and $k \in [\kin-1]$.  In the case $B(s)= \Id_q$,
    {\tt 3P-SPIDER} is a perturbed proximal-gradient algorithm (see
    e.g. \cite{atchade:fort:moulines:2017}); the convergence analysis
    below addresses the non convex case ($\lyap$ is not assumed to be
    convex). In the case $g=0$ and $\Sset = \rset^q$, {\tt 3P-SPIDER}
    is a Stochastic Approximation algorithm designed to find the roots
    of the preconditioned gradient $s \mapsto \mf(s) = -B^{-1}(s)
    \nabla \lyap(s)$. Applied with $g= 0$ or $g= \indic_{\mathcal{K}}$
    - the characteristic function of a closed convex set
    $\mathcal{K}$, {\tt 3P-SPIDER} is a variance reduced incremental
    EM algorithm (see
    \cite{fort:moulines:gach:2020,fort:moulines:SSP21:algorithm}, see
    also section~\ref{sec:logistic} for an application to the
    minimization of a penalized empirical loss).

    \section{Convergence in expectation and complexity bounds}
    \label{sec:complexity}
For ease of exposition (see \cite{fort:moulines:2021} for the general
case), it is assumed hereafter that
\begin{assumption}
  \label{hyp:local}
  $\pas_{t+1,0} =0$ and $\mathcal{E}_{t+1} =0$ for any $t \in
        [\kouter]^\star$. \\ The intractable functions $s \mapsto
        \mf_i(s)$ are defined as an integral with respect to (w.r.t.)
        a distribution $\pi_{i,s}$
\[
\mf_i(s) \eqdef \int_\Zset \mfint_i(z) \, \pi_{i,s}(\rmd z) \eqsp.
\]
They are approximated by a Monte Carlo sum:
\[
\mf_i(\hatS_{t,k-\ell}) \approx \hatmf_i^{t,k-\ell} \eqdef
\frac{1}{\nbrmc_{t,k+1}} \sum_{r=1}^{\nbrmc_{t,k+1}}
\mfint_i(Z_{r}^{i,t,k-\ell}),  \ell \in \{0,1\}
\]
where conditionally to the past of the algorithm $\F_{t,k}$, the
random variables $\{Z_r^{i,t,k-\ell}, r \geq 0 \}$ are independent and
identically distributed (i.i.d.) with distribution
$\pi_{i,\hatS_{t,k-\ell}}$.
\end{assumption}
Section~\ref{sec:logistic} provides an example of this setting. More
precisely, $\F_{t,k}$ is the filtration associated to the history of
the algorithm up to the outer loop $\# t$ and the inner loop $\# k$,
\begin{align*}
\F_{t,-1} & \eqdef \F_{t-1, \kin} \eqsp, \quad \F_{t,0} \eqdef
\F_{t,-1} \bigvee \sigma(\hatmf_i^{t,-1}, i \in [n]^\star) \eqsp,
\\ \F_{t,k} &\eqdef \F_{t,k-1} \bigvee \sigma\left( \batch_{t,k};
\hatmf_i^{t,k-1}, \hatmf_i^{t,k-2}, i \in \batch_{t,k} \right) \eqsp.
\end{align*}
 For
any $t \in [\kouter]^\star$ and $k \in [\kin-1]$, define
\[
\error_{t,k+1} \eqdef \frac{1}{\lbatch} \sum_{i \in \batch_{t,k+1}}
\left( \hatmf^{t,k}_i - \hatmf^{t,k-1}_i - \mf_i(\hatS_{t,k}) +
\mf_i(\hatS_{t,k-1}) \right) \eqsp;
\]
$\error_{t,k+1}$ corresponds to the errors when approximating the
quantities $\mf_i(\hatS_{t,k-\ell})$ for $\ell \in \{0,1\}$ at outer
loop $\# t$ and inner loop $\# (k+1)$. From standard computations on
i.i.d. random variables (the randomness being the selection of
the mini-batch $\batch_{t,k+1}$), we have
\begin{align}
& \PE\left[ \error_{t,k+1} \vert \F_{t,k} \right] = 0
  \eqsp, \label{eq:error:bias} \\ & \PE\left[ \|\error_{t,k+1} -
    \PE \left[ \error_{t,k+1} \vert \F_{t,k} \right] \|^2 \vert
    \F_{t,k}\right] \leq \frac{C_v}{ \lbatch \, \nbrmc_{t,k+1}}
  \eqsp, \label{eq:error:variance}
\end{align}
where
\[
C_v \eqdef 2 \sup_{s \in \Sset} n^{-1} \sum_{i=1}^n \int_\Zset
\|\mfint_i(z) - \mf_i(s) \|^2 \pi_{i,s}(\rmd z) \eqsp.
\]The result \eqref{eq:error:bias} claims
that the errors $\eta_{t,k+1}$ are unbiased, while
\eqref{eq:error:variance} is the control of its conditional variance -
which is a decreasing function of the batch size and the number of
points in the Monte Carlo sum. In \eqref{eq:error:variance}, the
control is uniform with respect to the past (the right hand side is
deterministic while the left hand side is random): this assumption can
be difficult to check when the optimization problem is not constrained
in order to ensure that the points $\hatS_{t,k}$ remain in a bounded
subset of $\rset^q$.

Theorem~\ref{theo:iid} provides an upper bound on a control in
expectation of the convergence of the algorithm. First, it controls
the difference of two successive values $\hatS_{t,k} - \hatS_{t,k-1}$;
then it controls the quantity
\[
\Delta_{t,k} \eqdef \frac{\|\Prox_{B(\hatS_{t,k}), \pas_{t,k} g} \hspace{-0.1cm}\left(
  \hatS_{t,k-1} + \pas_{t,k} \mf(\hatS_{t,k-1}) \right) \hspace{-0.1cm}
  - \hatS_{t,k-1} \|^2}{\pas^2_{t,k}}
\]
which is a kind of distance to the set of the solutions to
\eqref{eq:problem} (see \eqref{eq:fixedpoint}). When $B(s) = \Id_d$
and $g=0$, $\Delta_{t,k} = \|\mf(\hatS_{t,k})\|^2 = \|\nabla
\lyap(\hatS_{t,k})\|^2$. The proof of Theorem~\ref{theo:iid} is given
in \cite{fort:moulines:2021}.
\begin{theorem}
  \label{theo:iid}
  Assume A\ref{hyp:Prox}, A\ref{hyp:precond} and A\ref{hyp:local}. For
  any $t \in [\kouter]^\star$ and $k \in [\kin-1]$, set
  \[
\pas_{t,k} = \pas_\star\eqdef \frac{v_{\min}}{L_{\dot \lyap} + 2L
  v_\max \sqrt{\kin} / \sqrt{\lbatch}}.
\]
Denote by $(\tau, K)$ a uniform random variable on
$[\kouter]^\star\times [\kin]$, independent of the path
$\{\hatS_{t,k}, t \in [\kouter]^\star, k \in [\kin] \}$.  Then,
\begin{align*}
& \frac{v_{\min}^2}{2 \left( L_{\dot \lyap} + 2 L v_{\max} \sqrt{\kin}/
   \sqrt{\lbatch} \right)} \PE\left[\frac{\|\hatS_{\tau,K} -
     \hatS_{\tau,K-1}\|^2}{\pas_{\tau,K}^2} \right] \\
 & \leq
 \frac{1}{\kouter (1+\kin)}\left( \lyap(\hatS_\init) + g(\hatS_\init) -
 \min(\lyap + g) \right) \\ & + C_v \, \frac{v_\max}{2 L}
 \frac{1}{\sqrt{\kin \lbatch}} \PE\left[ \frac{\kin -
     K}{\nbrmc_{\tau,K+1}} \right] \eqsp.
\end{align*}
In addition
\begin{align*}
&\left( \frac{2}{v_{\min}} \left\{ L_{\dot \lyap} + 2L v_\max
  \sqrt{\frac{\kin}{\lbatch} } \right\}+ L \sqrt{\frac{\kin}{\lbatch}
  } \right)^{-1} \PE\left[ \Delta_{\tau,K} \right] \\ & \leq \left\{
  \frac{L_{\dot \lyap}}{L v_{\min}} + 2 \frac{v_\max}{v_{\min}}
  \sqrt{\frac{\kin}{\lbatch} }
  \right\}^{-1} \hspace{-0.2cm}\left(\frac{1}{L}+ \left(
  \frac{v_\max}{v_{\min}}\right)^2\pas_\star
  \sqrt{\frac{\kin}{\lbatch}} \right) \cdots \\ & \qquad \qquad \times
  \PE\left[ \frac{\| \hatS_{\tau,K} -
      \hatS_{\tau,K-1}\|^2}{\pas^2_{\tau,K}} \right] \\ & \qquad +
  \left(\frac{v_\max}{v_{\min}}\right)^2 \frac{C_v}{L}
  \frac{1}{\sqrt{\lbatch \kin}} \PE\left[
    \frac{\kin-K}{\nbrmc_{\tau,K}} \right] \eqsp.
\end{align*}
\end{theorem}
In Theorem~\ref{theo:iid}, the expectations are w.r.t. the stochastic
path of the algorithm $\{\hatS_{t,k}\}$ and to the randomness of the
times $(\tau, K)$. This theorem provides a control of the errors when
the algorithm is stopped at some random time $(\tau, K)$. Such a control is
classical in the non-convex case to show non-asymptotic convergence of
stochastic gradient methods to a stationary point
\cite{ghadimi:lan:2013}: it consists in fixing a maximal number of
iterations (here set to $\kouter \times \kin$) of the algorithm, and
draw at random, prior the run of the algorithm, a stopping time
$(\tau, K)$ .

Let us discuss the complexity bounds in the case $m_{t,k} = m$ for any
$t \in [\kouter]^\star$ and $k \in [\kin]^\star$. The total number of
proximal calls is equal to $ \mathcal{N}_{P} \eqdef \kouter ( \kin
+1)$. The total number of approximations of the functions $\mf_i$
is equal to $\mathcal{N}_{A} \eqdef \kouter (n+ 2 \lbatch \kin)$. The
total number of Monte Carlo draws is $\mathcal{N}_{MC} \eqdef m
\mathcal{N}_A$. Let us fix $\epsilon >0$. Among the values
$\mathcal{K}(n,\epsilon)$ of the positive integers $(\kouter, \kin,
\lbatch, m) \in \nset^4$ which guarantee that $\epsilon$-stationarity
is reached i.e.
\[
\PE\left[ \Delta_{\tau,K} \right] \leq \epsilon \eqsp,
\]
the complexity bounds in terms of proximal calls are defined as
$\mathcal{K}_{P} \eqdef \min_{\mathcal{K}(n,\epsilon)} \mathcal{N}_P$;
similarly, we define the complexity in terms of $\mf_i's$
approximations, and in terms of Monte Carlo draws.  By choosing
$\lbatch = \kin =\sqrt{n}$, $\kouter = 1/(\sqrt{n} \epsilon)$ and $m=
1/ \epsilon$ , it is easily seen from Theorem~\ref{theo:iid} that
$\PE\left[ \Delta_{\tau,K} \right] = O(\epsilon)$ and
\[
\mathcal{K}_P =O(\epsilon^{-1}) \eqsp, \quad \mathcal{K}_{A} =
O(\sqrt{n} \epsilon^{-1}) \eqsp, \quad \mathcal{K}_{MC} = O(\sqrt{n}
\epsilon^{-2})\eqsp.
\]
When $B(s) =\Id_q$, $g=0$ and the gradient functions $\mf_i$'s can be
computed exactly, the state of the art complexity of variance-reduced
gradient algorithm in terms of total number of computations of these
gradient functions is $O(\sqrt{n} \epsilon^{-1})$
\cite{wang:etal:nips:2019}.  This bound is also reached by the
variance reduced incremental EM named {\tt SPIDER-EM}, which
corresponds to the case $g=0$, $B(s) \neq \Id_d$, and the
preconditioned gradient functions $\mf_i$'s are explicit (see
\cite{fort:moulines:wai:2020}). Our complexity $\mathcal{K}_A$ reaches
this optimal value: in that sense, {\tt 3P-SPIDER} is optimal since
the bound $O(\sqrt{n} \epsilon^{-1})$ is reached despite the
introduction of a proximal operator and the approximations of the
preconditioned gradient functions $\mf_i$'. To reach this optimal
bound, the Monte Carlo complexity is $O(\sqrt{n} \epsilon^{-2})$.

\section{Application: Inference in the Logistic Regression Model}
\label{sec:logistic}
We illustrate the convergence of {\tt 3P-SPIDER} applied to inference
in the following logistic regression model. Given $n$ covariate
vectors $\{X_i, i \in [n]^\star \}$ in $\rset^d$ and $\param \in
\rset^d$, the $\{-1, 1\}$-valued binary observations $\{Y_i, i \in
     [n]^\star \}$ are assumed independent with success probability
     $\PP_\param(Y_i = 1)$ equal to
\[
\frac{1}{\sigma^d \sqrt{2 \pi}^d}\int_{\rset^d} \left(1 + \exp(- \pscal{X_i}{z_i}) \right)^{-1} \,
\exp\left( - \frac{\|z_i - \param\|^2}{2 \sigma^2}\right) \rmd z_i \eqsp.
\]
This model corresponds to a predictor $Z_i$ for each individual $\# i$
and these predictors $Z_j, j \in [n]^\star$, are independent with
distribution $\mathcal{N}_d(\param, \sigma^2 \Id_d)$. It is assumed
that $\sigma^2$ is known; the unknown parameter $\param$ is learnt by
minimization of the penalized normalized negative log-likelihood
$\param \mapsto F(\param)$, with penalty term $\pen(\param) \eqdef
\tau \|\param\|^2$ for some $\tau >0$. $F$ is equal to (see
\cite{fort:moulines:SSP21:algorithm})
\begin{multline*}
\param \mapsto -\frac{1}{n} \sum_{i=1}^n \log \int_\rset \left( 1 +
\exp\left(-Y_i \|X_i\| z \right) \right)^{-1} \exp\left(
\pscal{\s_i(z)}{\param}\right) \\ \times \, \exp( - z^2/(2\sigma^2))
\rmd z + \R(\param)
\end{multline*}
where $\s_i(z) \eqdef z X_i / (\|X_i\| \sigma^2)$, $\R(\param) \eqdef
(1/2) \param^T \Omega^{-1} \param$ and $\Omega \eqdef \left(
\frac{1}{\sigma^2 n} \sum_{i=1}^n \frac{X_iX_i^T}{\|X_i\|^2} + 2 \tau
\Id_d \right)^{-1}$. The minimization of this criterion by a EM
algorithm can be solved equivalently in the expectation space in order
to minimize $\lyap: s \mapsto F(\Omega s)$ (see
e.g. \cite{Delyon:lavielle:moulines:1999,fort:moulines:SSP21:algorithm}). In
that case, EM finds the roots on $\rset^d$ of
\[
s \mapsto \mf(s) \eqdef \frac{1}{n} \sum_{i=1}^n \int_\rset \s_i(z)
p_i(z;\Omega s) \rmd z -s \eqsp;
\]
$z \mapsto p_i(z;\param)$ is the probability density function
proportional to
\[
z \mapsto \left( 1 + \exp\left(-Y_i \|X_i\| z \right) \right)^{-1}
\exp\left( \pscal{\s_i(z)}{\param} - z^2/(2\sigma^2) \right).
\]
We have $\nabla \lyap(s) \eqdef - \Omega \mf(s)$ for any $s \in
\rset^d$ (see \cite{fort:moulines:SSP21:algorithm}).  Furthermore,
upon noting that $\PP_\param(Y_i = y_i) \leq 1$, it can be shown that
the minima of $F$ are in the set $\{\param \in \rset^d: \tau
\|\param\|^2 \leq \ln 4 \}$, which implies that EM in the expectation
space will find the roots of $\mf$ in $\mathcal{K} \eqdef \{s \in
\rset^d: s^T \Omega s \leq \ln 4 /(\tau \lambda_{\min}) \}$ where
$\lambda_{\min}$ is the positive minimal eigenvalue of
$\Omega$. Therefore, we set $g$ equal to the characteristic function
of $\mathcal{K}$; with such a definition of $\mathcal{K}$, the
associated weighted proximal is explicit. \\ The data set is built
from the MNIST data set, as described in \cite[Section
  3]{fort:moulines:SSP21:algorithm}: $n = 24 \, 989$ and $d=51$. {\tt
  SPIDER-EM} is run with $\sigma^2 = 0.1$, $\tau = 1$, $\kouter = 20$,
$\kin= \lceil \sqrt{n}/10 \rceil =16$, $\lbatch = \lceil 10 \sqrt{n}
\rceil$, $\mathcal{E}_t= 0$, $\pas_{t,0} = 0$, $\pas_{t,k} = 0.1$ and
$m_{t,k} = 2 \lceil \sqrt{n} \rceil $ until the outer loop $\# 9$ and
then $m_{t,k} = 10 \lceil \sqrt{n} \rceil $.

\begin{figure}[h] \centering
\begin{minipage}[b]{0.48\linewidth}
 \includegraphics[width=4.6cm]{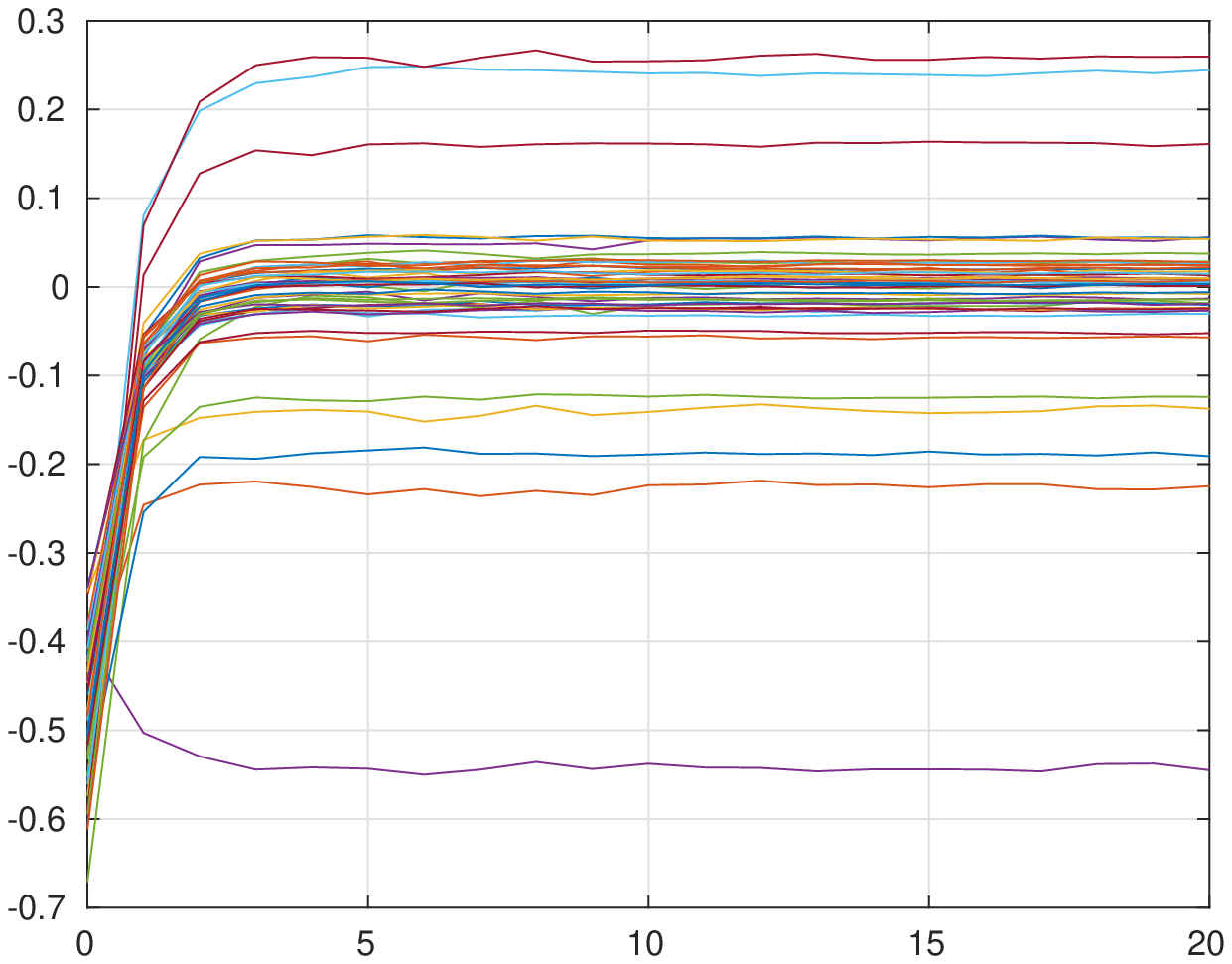} \\
 \includegraphics[width=4.6cm]{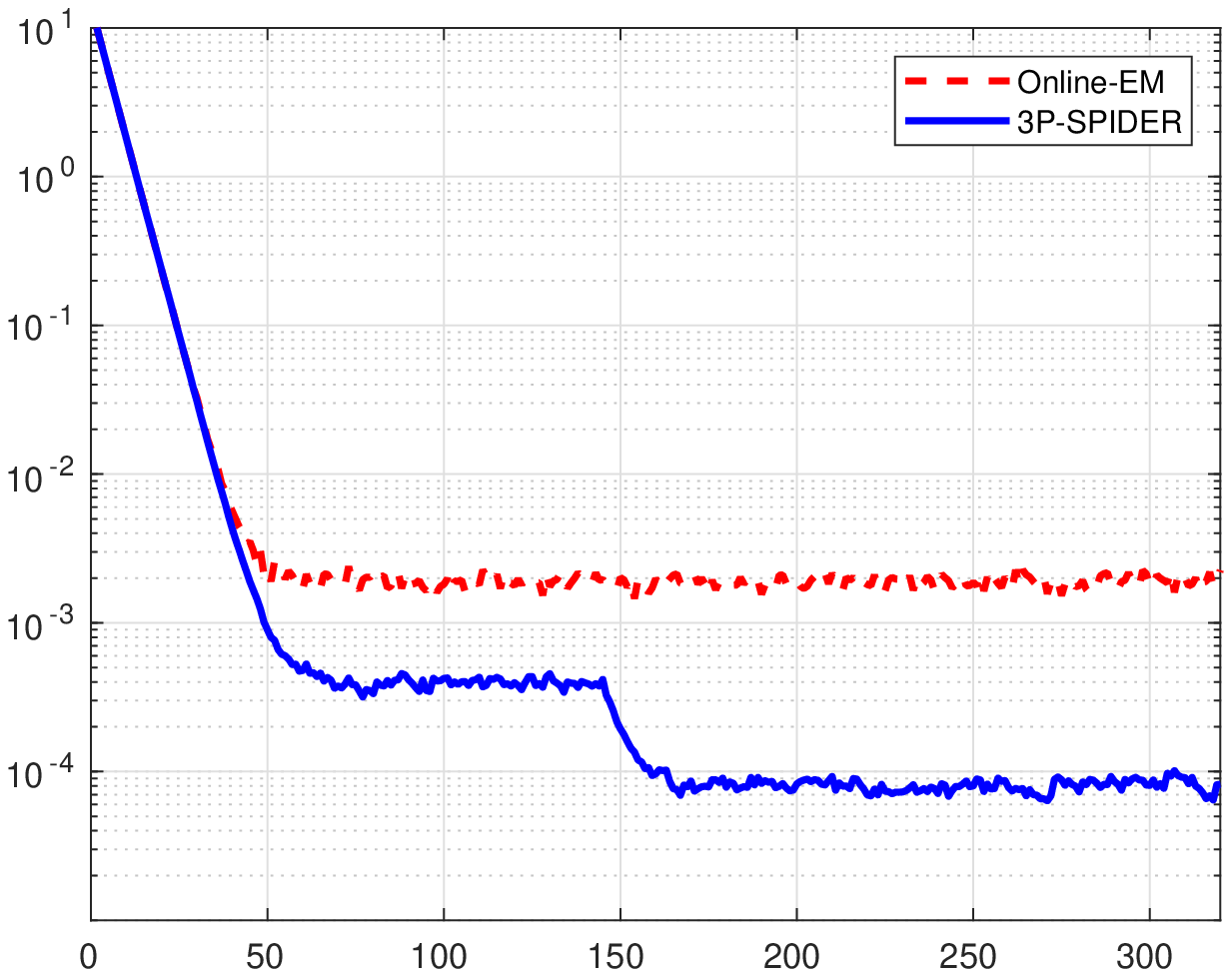}
 \end{minipage}
\hfill
 \begin{minipage}[b]{0.48\linewidth}
  \includegraphics[width=4.6cm]{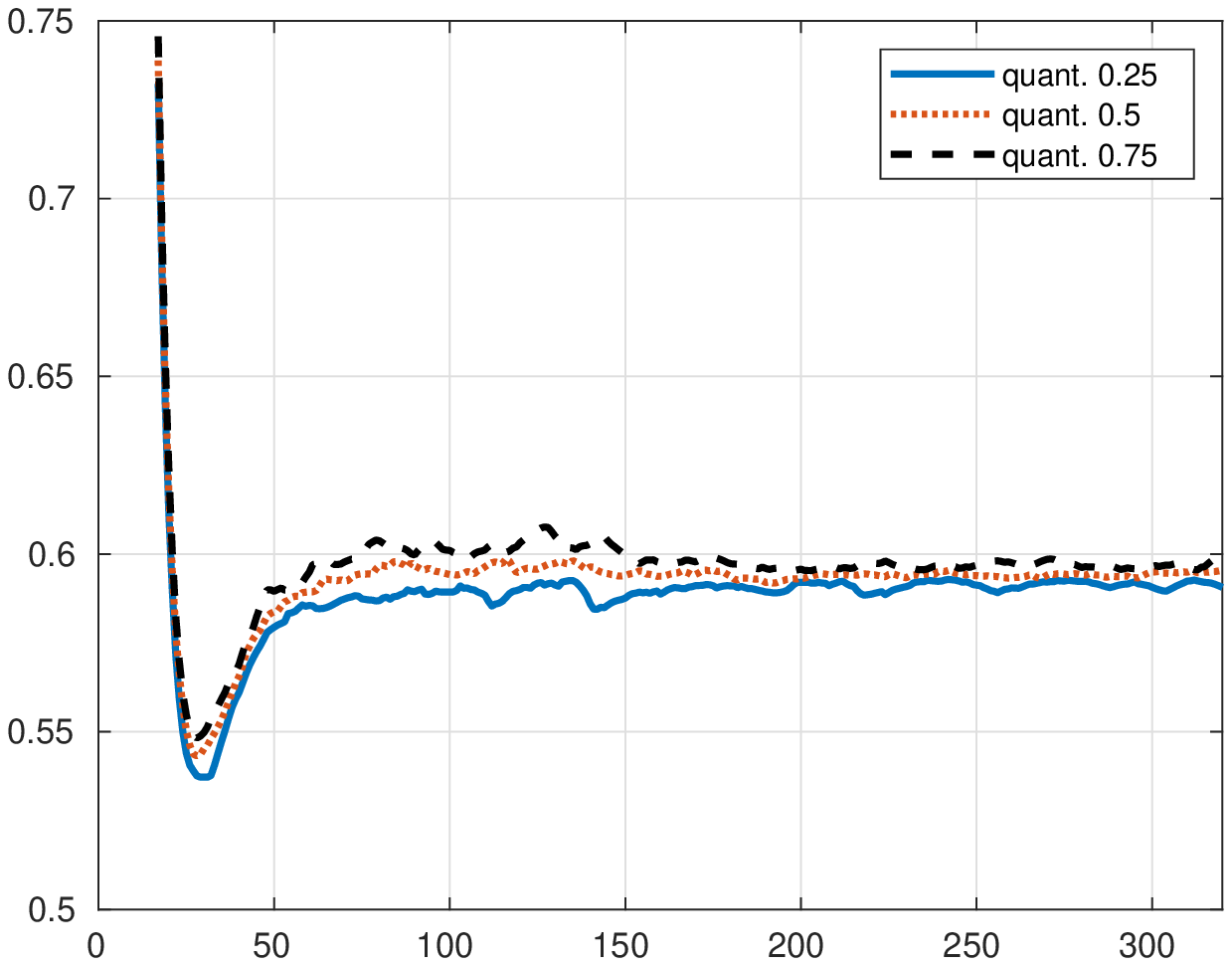} \\
  \includegraphics[width=4.6cm]{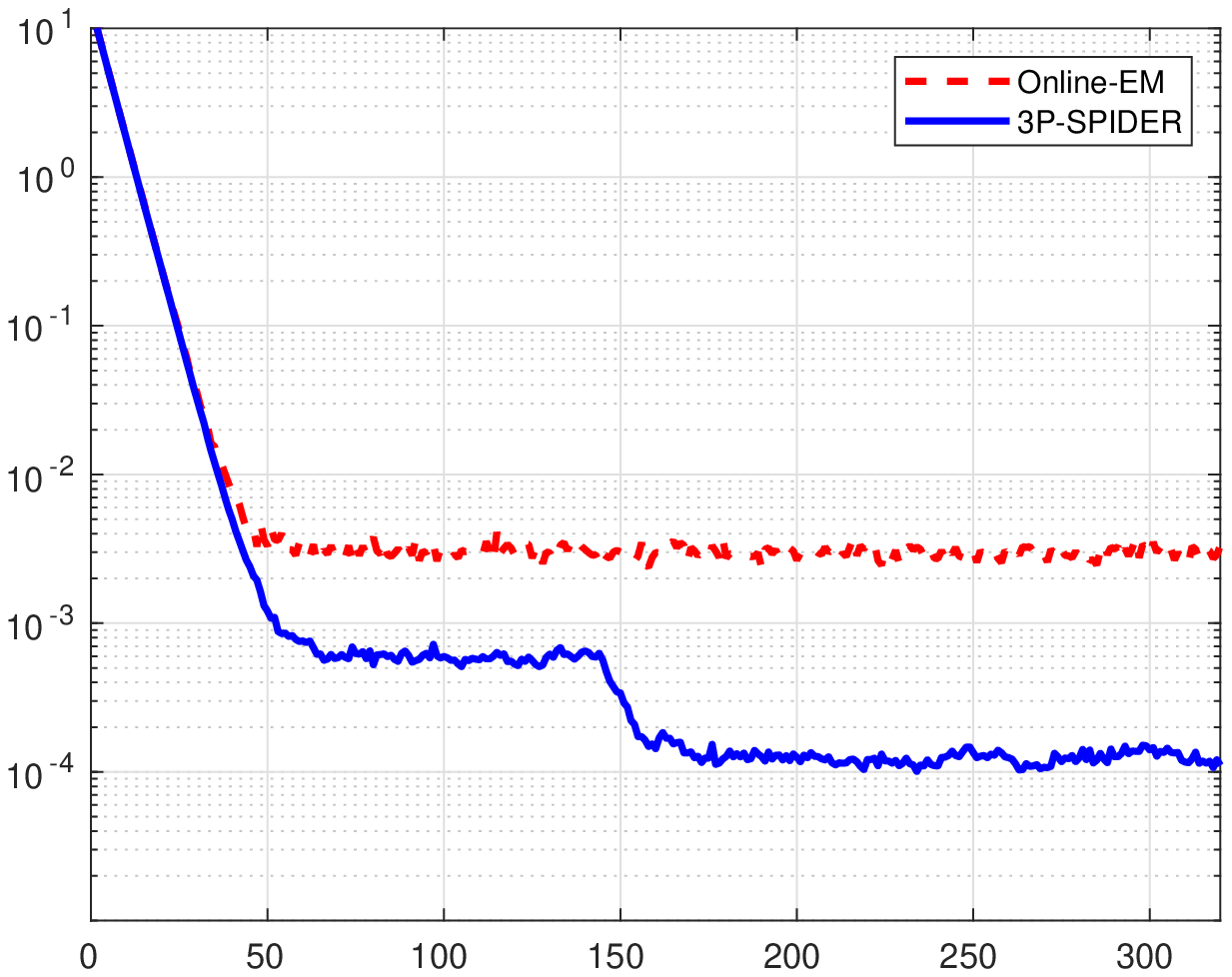}
\end{minipage}
 \caption{[(a) top left] Sequence $\{\hatS_{t,\kin}, t \in
   [\kouter]^\star\}$ [(b) top right] Quantiles of $\|\hatS_{t,k}\|^2$
   [(c) bottom left] Quantile $0.25$ of $\widehat{\Delta}_{t,k} $ [(d)
     bottom right] Quantile $0.75$ of $\widehat{\Delta}_{t,k} $}
\label{fig:plot}
\end{figure}

On Figure~\ref{fig:plot}(a), the $51$ components of the sequence
$\{\hatS_{t,\kin}, t \in [\kouter]^\star\}$ are displayed vs the index
of the outer loop $t$. The convergence can be observed.

On Figure~\ref{fig:plot}(b), we display the quantiles $0.25$, $0.5$
and $0.75$ of the squared norm $\|\hatS_{t,k}\|^2$ as a function of
the cumulated number of inner loops; these quantiles are computed over
$25$ independent runs of {\tt 3P-SPIDER}. Here again, the convergence
and the stability of the path over the independent runs can be
observed.

Finally, Figures~\ref{fig:plot}(c,d) display the quantiles $0.25$ and
$0.75$ of $ \widehat{\Delta}_{t,k} \eqdef \| \hatS_{t,k} -
\hatS_{t,k-1}\|^2 / \pas_{t,k}^2$ as a function of the cumulated
number of inner loops; these quantiles are estimated over $25$
independent runs of {\tt 3P-SPIDER}. We observe the gain when
increasing the number of Monte Carlo points in order to reduce the
fluctuations of the approximations of the $\mf_i$'s; see \cite[Section
  3]{fort:moulines:SSP21:algorithm} for a detailed study of the design
parameters of {\tt 3P-SPIDER}. To illustrate the benefit of the
variance-reduction step in {\tt 3P-SPIDER}, we also run {\tt
  Prox-Online-EM} with $\lbatch = \lceil 10 \sqrt{n} \rceil$ and
$m_{t,k} = 2 \lceil \sqrt{n} \rceil$.  {\tt Prox-Online-EM}
corresponds to {\tt Online-EM} combined with a proximal step i.e. a
proximal-preconditioned gradient algorithm. The quantiles $0.25$ and
$0.75$ of $\| \hatS_{t} - \hatS_{t-1}\|^2 / \pas_{t}^2$ are displayed
on Figures~\ref{fig:plot}(c,d) as a function of the number of
iterations $t$. It illustrates that {\tt 3P-SPIDER}, as a proximal
variance-reduced preconditioned gradient method, clearly improves on
{\tt Online-EM}.

\end{document}